\documentclass[aps,pra,twocolumn]{revtex4}
\usepackage{psfig}
\newcommand{\beq}{\begin{equation}}
\newcommand{\eeq}{\end{equation}}
\newcommand{\beqa}{\begin{eqnarray}}
\newcommand{\eeqa}{\end{eqnarray}}
\newcommand{\beqar}{\begin{eqnarray*}}
\newcommand{\eeqar}{\end{eqnarray*}}
\newcommand{\tr}{{\rm tr}}

\newtheorem{theorem}{Theorem}

\def \s {\,\,\,\,}
\def \la {\langle}
\def \ra {\rangle}
\def \up {\uparrow_z}
\def \down {\downarrow_z}
\def \r {\rho}

\newcommand{\proj}[1]{\ket{#1}\bra{#1}}
\newcommand{\bra}[1]{\langle #1 |}
\newcommand{\ket}[1]{| #1 \rangle}

\newcommand{\qed}{\quad\blackslug\lower
8.5pt\null}
\newcommand{\blackslug}{\hbox{\hskip 1pt \vrule
width 4pt height 8pt
depth 1.5pt \hskip 1pt}}
\newcommand{\E}[1]{{\cal E}( #1 )}
\def \U {U}

\def \rs {\varrho}
\def \M {M}
\def \N {N}
\def \so {S_{\cal{E}}}
\def \C {C}
\def \en {\zeta}
\def \x {X}
\def \singlet {\psi^{+}}
\newcommand{\sing}[1] {\psi_#1}
\def \A {A(t_2,t_1)}

\begin{document}

\title{A probabilistic and information
theoretic interpretation
of quantum evolutions
}

\author{J. Oppenheim}
\email{jono@damtp.cam.ac.uk} \affiliation{
Dept. of Applied Mathematics and Theoretical Physics, University of
Cambridge, Cambridge, U.K.}
\affiliation{Racah Institute of Theoretical Physics,
Hebrew University of Jerusalem, Givat Ram, Jerusalem 91904, Israel}

\author{ B. Reznik }
\email{reznik@post.tau.ac.il} \affiliation{ Department of
Physics
and Astronomy, Beverly and Raymond Sackler Faculty of Exact Sciences,
 Tel-Aviv University, Tel Aviv 69978, Israel.
       }

\date{\today}
\begin{abstract}
In quantum mechanics, outcomes of measurements
on a state
have a probabilistic interpretation while the
evolution of the
state is treated deterministically.  Here we
show that one can
also treat the evolution as being probabilistic
in nature
and one can measure `which unitary' happened.
Likewise, one
can give an information-theoretic
interpretation to evolutions
by defining the entropy of a completely
positive map.  This
entropy gives the rate at which the
informational content of the evolution
can be compressed.  One cannot
compress this
information and still have the evolution act on
an unknown state, but we demonstrate a general scheme to 
do so probabilistically.
This allows one to generalize super-dense
coding to the sending
of quantum information.
One can also define the ``interaction-entanglement'' of a
unitary, and concentrate
this entanglement.
\end{abstract}

\pacs{}


\maketitle

\section{Introduction}

An isolated system is represented in Quantum Mechanics by a state
vector that conveys statistic predictions for measurement outcomes
and manifests phenomena such as superpositions, and entanglement.
The temporal evolution law of the state is determined by the
unitary operator $U=\exp -iHt/\hbar$ where the Hamiltonian $H$ is
dictated either by external classical potentials and/or universal
inter-particle/fields interactions. Therefore, while  the state
vector manifests the non-deterministic features of Quantum
Mechanics, the temporal evolution law of an isolated system is
regarded as fully deterministic. This asymmetry between the properties of
states and evolutions is also maintained  within the framework of quantum
information theory wherein the information is carried by the state alone.

In this work we examine the consequence of measurements of the
transformation law and suggest that the above restricted view of
quantum evolutions can be extended even within the conventional
framework of quantum mechanics and quantum information. We find
that features such as superposition of unitary evolutions,
collapse to a certain evolution and a corresponding probability
law, can in fact be attributed in a natural fashion to unitary
evolutions, as well as to the more general case of non-unitary
evolutions that can be described by completely positive (CP) trace
preserving linear maps. Particularly, we show that a measurement
of `which evolution occurred' during  a certain time interval
`collapses' the quantum evolution to a particular evolution
with a probability given by a simple extension of the ordinary
probability law. Our results provide an operational meaning to a
formal correspondence between states and operations introduced by
Jamiolkowski \cite{jam-72}.

Although the present work is aimed at extending concepts ascribed
to states into the domain of evolutions, it also has applications
to quantum computation in that we present methods which can be
used to monitor the interactions of a quantum device without
changing the physical set-up of the device.

Next, we turn to the question of whether operations have
informational content in a manner analogous to quantum states.  We
find that one can assign a state independent entropy to an
arbitrary completely positive (CP) map, and that this entropy
gives the rate at which the informational contents of the map can
be compressed. Here, the information content refers to the
sequence of Kraus operators \cite{kraus1}
 used to implement the CP map, although other
interpretations are possible. This can be considered as the
equivalent of Schumacher's noiseless coding theorem for
operations. A different interpretation of compression and storage
of unitaries was given in \cite{durcirac-ops} where for a specific
known ensemble of phase gates it was shown how to store them
efficiently.  This can be thought of as storage
of an ensemble of evolutions \cite{VMC-2001}. 
Here, our compression rate is ensemble independent and generic,
akin to compression of a source emitting quantum states.

We invoke a no-go theorem \cite{nielsonchuang-universal,VMC-2001} 
for programmable quantum gates, and storage of 
unitaries to show that
one cannot have a compressed evolution act on an unknown state, and
still preserve its informational content.  This is true even
if one only demands approximate fidelity.
This is because there
are an infinite number of ways one can implement a given CP map 
(there are in a sense, an infinite number of {\it evolution ensembles},
which are unknown).  We show however, a generic scheme to probabilistically
act the evolution on an unknown state. 

We then generalize super-dense coding to the
sending of quantum information contained in unitaries.
This has certain cryptographic implementations which
we briefly explore.

Finally, we turn to the notion of entanglement
of a unitary.
A number of authors have used the formal
correspondence between
states and operations to investigate the
entangling capabilities
of unitary operations \cite{clp-2000},(e.g.
\cite{cdkl-2000,durcirac-ops,zanardi-etal,zanardi-2001,wang-zanardi}).
The present framework suggests the notion of
{\it interaction-entanglement}
of a unitary acting on systems, and we show
that this entanglement
can be concentrated in a manner analogous to
the concentration
of states into pure entanglement.  We conclude
with some remarks
on the interpretational issues involved with
the measurement
of evolutions.

\section{A probabilistic interpretation of unitaries}

Let us first provide an operational meaning to
the measurement of
unitaries.
We consider a system with an $N$-dimensional
Hilbert-space
whose state evolves in time according to
\beq
|\psi(t_1)\ra\to |\psi(t_2)\ra = U(t_2-t_1)
|\psi(t_1)\ra
\eeq

It is know that for any $N$ there exists a basis of $N^2$
orthogonal unitary operators \cite{werner-alltel}, where
orthogonality is defined with respect to the trace inner-product
$U\cdot V \equiv {\rm tr} U^\dagger V$. Thus the unitary time
evolution operator can be decomposed with respect to an orthogonal
basis $U_\alpha$
\beq\label{superposition} U =
\sum_{\alpha=0}^{N^2-1} C_{\alpha} U_{\alpha} \eeq
where $U_\alpha
\cdot U_\beta = N\delta_{\alpha\beta}$ and the complex amplitudes
are given by $C_\alpha = (1/N) U_\alpha \cdot U(t_2-t_1)$.
The
converse of the above statement is not true. A superposition of
unitary operators with arbitrary amplitudes generally does not
give rise to a unitary. The operators space contains non-unitary
operators which can be also spanned by a unitary basis.

Can the formal expansion (\ref{superposition})
be given a general physical interpretation?
It has been shown that under certain conditions,
a superposition of unitary evolutions that gives rise to
another unitary, can be produced by
post-selecting an ancillary system that interacts weakly
with our system.
In the present work we propose another approach.
We shall show that for any chosen basis,
we can measure which unitary evolution $U_\alpha$ the
system evolved under. The  outcome of such a measurement 
has probability  ${\rm Prob}(U_{\alpha_0})
= |C_{\alpha_0}|^2$.
More formally:

\smallskip
\noindent
1) {\bf Observables and Eigenvalues}.
{\em To each orthogonal basis of unitary
operators, we can find
an observable $A(t_2,t_1)$ that assigns to
each unitary $U_\alpha$ a distinct real
eigenvalue $\lambda_\alpha$
through the eigenvalue equation}
\beq \label{eigenvalues}
T:A(t_2;t_1) U_{\alpha} = \lambda_{\alpha}
U_{\alpha}
\eeq

\smallskip

The operator  $A(t_2,t_1)$ describes temporal correlations.
It is constructed as a linear combination of  bilinear products of
operators at each of the two instances $t_2$ and $t_1$.
The symbol $T:$ denotes  temporal ordering. For instance if
$A(t_2;t_1)= \alpha A(t_2)B(t_1) + \beta C(t_2)D(t_1)$ then
$T:A(t_2;t_1)U_\mu(t_2-t_1) = \alpha AU_\mu B + \beta CU_\mu D$.
Since $\lambda_\mu =\alpha AU_\mu B U_\mu^\dagger  + \beta CU_\mu
D U_\mu^\dagger$
is a constant, the operator $A(t_2,t_1)$ describes constant of
motion with respect to each of the basis elements.
The eigenvalues are thus state independent.

 \smallskip
\noindent 2) {\bf Probability Law and Measurements.} {\em The
outcome of a measurement of $A(t_2,t_1)$ is one of eigenvalues
$\lambda_{\alpha_0}$ with a probability given by }
\beq
{\rm Prob}(\lambda_{\alpha_0}) = |C_{\alpha_0}|^2
\eeq

 \smallskip

\noindent
and
\newline

3) {\bf Reduction of $U$ (Collapse).}

{\em a  measurement with an outcome
$\lambda_{\mu_0}$ leads to a
collapse (effectively or truly  depending to
the readers preferred interpretation)
of the superposition
(\ref{superposition}) according to }
\beq
U(t_2-t_1)|\psi\ra \to  U_{\mu_0}|\psi\ra
\eeq

 \smallskip

We interpret 1-3 as specifying criteria for a measurement
that detects which particular transformation
$U_\alpha$ in the superposition
(\ref{superposition}) has been realized on the system
with a priori probability  ${\rm Prob}(U_{\alpha_0}) = |C_{\alpha_0}|^2$.

It should be emphasized that 1-3 are
independent of the initial state of the system and hence can be
interpreted as a measurement of a term in the superposition
(\ref{superposition}).
The initial state of the system is here
arbitrary, hence includes the
case of a unitary acting (locally) on a part of
an entangled state.
As a consequence, the present measurement of
the unitary transformation
{\em does not reduce} the entanglement of the system.

Although here we will show an operational correspondence between
the measurement of states and the measurement of unitaries, it
must be emphasized that there are important differences.  One
interesting result is that one can distinguish between two
unitaries which are not orthogonal \cite{acin-unitaries}.

We now proceed to prove the above three statements.
To begin with, we consider some simple
properties of a general given basis of of orthogonal unitary
operators  $\lbrace U_\alpha \rbrace$ Eq.
(\ref{superposition}).
Clearly the set $\lbrace 1, U'_i;\
i=1,..,d-1\rbrace$
where $U'_\alpha = U_0^\dagger U_i$ is also
orthogonal.
Since $U_i$ are traceless orthogonal operators,
all sets of unitary orthogonal basis
can by expressed as a product of an arbitrary
fixed unitary $U_0$ with
some traceless unitary orthogonal basis.
As a consequence of this structure we have that
\beq
U^\dagger U = \sum_\alpha |C_\alpha|^2  +
\sum_{\alpha\beta}
C^*_\alpha C_\beta U^\dagger_\alpha U_\beta = 1
\eeq
but from the nature of the matrices we have
that
\beq
\sum_\alpha |C_\alpha|^2 =1
\eeq

We explicitly consider the $N=2$ case --
generalizing our results to higher
dimensional Hilbert spaces (including the
infinite dimensional case) is straightforward and described in the Appendix.
The general structure of the basis is given be
\beq\label{2dset}
U_\alpha =  U_0 \sigma_\alpha
\eeq
where and $\sigma_\alpha =(1,\sigma_i)$
with  $i=x,y,z$.

The observables corresponding to the measurement of $U_\alpha$ can then be chosen as
\beq
A_i(t_2,t_1) = [U_0\sigma_i
U_0^\dagger]_{t_2}][\sigma_i]_{t_1}
\eeq
Replacing into (\ref{eigenvalues})
\beq
T:A_i(t_2,t_1)U_\alpha = [U_0\sigma_i
U_0^\dagger] U_\alpha [\sigma_i]=
\lambda_{i\alpha}
 U_\alpha
\eeq
and using (\ref{2dset}), we get
\beqa
\lambda_{i\alpha} &=&  [U_0\sigma_i
U_0^\dagger]U_\alpha \sigma_i
U_\mu^\dagger \\
&=&U_0\sigma_i U_0^\dagger U_0\sigma_\alpha
\sigma_i \sigma_\alpha
U_0^\dagger \\
&=&\sigma_i\sigma_\alpha \sigma_i \sigma_\alpha
= \pm 1
\eeqa
Since we need to resolve
between four basis elements, it is sufficient
to consider a
pair of operators, say,  $A_i(t_2;t_1)$ with
$i=z,x$.

Next demonstrate (2) by explicit
construction of a measurement.
One possibility is to have the unitary act on
half of a maximally
entangled state.  Each
orthogonal unitary
in a basis of unitaries would then produce an
orthogonal maximally
entangled state and one could then perform a
measurement on the
state to determine which unitary acted.  This
has the disadvantage
that one cannot use this method for an
evolution acting on a
particular physical system in an unknown state.  
We therefor
propose to
observe the operator $A(t_2,t_1)$ by coupling
twice with the system in a manner which
preserves the
 state.  A method for measuring sums of
operators
as $\sigma(t_2)+\sigma(t_1)$
has been suggested  \cite{aharonov-albert} and used to
demonstrate
teleportation \cite{vaidman-teleportation}.
We employ a similar method using a pair of
ancillary
two-level particles taken initially in the
state
$(|0\ra+|1\ra)(\tilde0\ra+|\tilde1\ra)/2$. We
assume a vanishing free
Hamiltonian for the ancillary particles.

The ancilla and the system then interact twice,
first at $t=t_1$ and then at $t=t_2$, and then
the ancilla is measured.
To specify the interaction between the system
and ancilla
we
define
\beq
V_i= |0\ra\la 0| + |1\ra\la1|\sigma_i
\eeq
where $\sigma_i$ acts on the system, and
similarly we denote by
$\tilde V_i$ the same interaction between the
system and the
second ancilla.
We further assume that the interactions are
nearly impulsive: the
duration
$\Delta t$  required to apply  $V_i \tilde V_i$
is much shorter than
$t_2-t_1$, hence the correction due to the free
evolution can be
neglected while we apply the interactions.

We now apply the following sequence
\beq
(U_0\tilde V_x V_z U_0^\dagger)
U(t_2-t_1)(\tilde V_xV_z)
\eeq
The measurement interaction acted twice at $t=t_1$
and $t=t_2$, while
at intermediate
times the system evolves freely.
The resulting total state  becomes
\beq
{1\over2}\sum_\mu C_\mu
(|\tilde0\ra +
\lambda_{x\mu}|\tilde1\ra)(|0\ra +
\lambda_{z\mu}
U_\mu |\psi\ra \\
\eeq
Finally, using the notation $|\mu\ra =\lbrace
\up\up,
\up\down,\down\down,
\down\up\rbrace$ where  $\up,\down = (|0\ra\pm
|1\ra)/\sqrt2$, the
final total state of the system and the two spins
and the effect of the measurement can be
expressed as
\beq
\sum_{\mu=0}^3 C_\mu|\mu\ra U_{\mu}|\psi\ra \to
U_{\mu_0}|\psi\ra
\eeq
thus demonstrating the notion of collapse (3).
One could also interpret the above as instead being a
collapse induced by our interaction.

We have used here the standard probability
interpretation with
respect to
a measurement of the final ancillary basis
$|\mu\ra$.
Since the probability to find $\mu_0$ is given
by $|C_{\mu_0}|^2$,
this demonstrates (3) and (4) for the present
two-dimensional
case.  One can verify that the above procedure
effectively moves
the information contained in the state onto the
ancilla, while
having the unitary act on half a maximally
entangled state.  The information of the state
(with the action of the unitary)
is then transferred back from the ancilla to the original
system.
However, the physical particle that is the
system is not actually
swapped, allowing one to use such a measurement
without changing
the particular system.  For example, one could
use this to detect
the noise in an ion-trap quantum computer while
still preserving the
information of the state and the set-up of the
experiment.  The measurement procedure gives
a generic way to transfer the state of a system onto
another system without performing a physical swap.

An important point is that the measurement of
`which unitary' is
independent of the state that the unitary acts
on.  This allows
one to distinguish between unitaries which for
certain states
would not lead to orthogonal outcomes.

To exemplify our result, consider the evolution of a spin in a
magnetic field with $U= \exp(-iB\sigma_z t)= cos(Bt)1 -i\sin(Bt)
\sigma_z$. If we select to measure in the basis $(U, U\sigma_x,
U\sigma_y, U\sigma_z)$, we will find  ${\rm Prob}(U) =1$ and ${\rm
Prob}(U\sigma_i)=0$. Therefore, in this case we verified with
certainty that the evolution is $U(t)$ without causing any
disturbance. On the other hand, if we choose to measure in the
basis $(1,\sigma_i)$, we will reduce the evolution to $1|\Psi\ra$
with probability $\cos^2(Bt)$, or to $\sigma_z|\psi\ra$ with
probability $\sin^2(Bt)$. More generally, in a $d$-dimensional
space, we can distinguish with certainty between $d^2$ orthogonal
unitary operators.

What is more, we are able to distinguish
between unitaries
which do not themselves commute.
This is
because each element of the basis gives
orthogonal outcomes on
maximally entangled states.  There is however
an uncertainty principle
between different possible measurements of
`which unitary' given
by the uncertainty between two operators $\A$
and $A'(t_2,t_1)$.  I.e.
\beq
[\A, A'(t_2,t_1)]\neq 0
\eeq

Finally, we comment that above procedure can be extended to
non-unitary orthogonal operators, which may be also used as a
basis. Such a non-unitary basis can be obtained by the
transformation $A_\mu= \sum_\nu K_{\mu\nu} U_\nu$, where $K$ is a
$N^2\times N^2$ dimensional unitary matrix. The operators $A_\mu$
are generally not unitary, but are orthogonal with respect to the
trace inner product. Thus, we can distinguish between the elements
$A_\mu$ using the procedure used above.

\section{An information theoretic interpretation of evolution}

Having shown that the correspondence between
unitaries and states
has an operational meaning in terms of
probability amplitudes,
we now turn to the question of whether there is
an information
theoretic interpretation to unitary operations.
An informational
interpretation of quantum states was given, by Schumacher's
noiseless coding theorem \cite{Schumacher1995}
(c.f. \cite{PetzMosonyi,HiaiPetz91}).
We will now see that a similar interpretation
can be given
to unitary operations.
Instead of considering a pure unitary, we
consider an arbitrary
completely positive (CP) map $\E{\r}$.  We will
see that one can define an entropy for the CP
map which only
depends on the map, and not on how it is
implemented, nor on what
state it acts, and that
this has an interpretation of the rate at which
the informational
contents of the map can be compressed.  It is
also equal to
the maximum classical information which the map
can transfer.
The entropy production that a CP map produces
on particular
states was considered in \cite{schumacker-96}.
We will further prove
two theorems showing that while the information
can be stored
and compressed, it is impossible to later act
it on an unknown
state, or even a known state chosen after the
information has
been stored.

We start by showing that the interpretation of unitaries described in
the previous section, can be extended to other positive operators.
Namely, we can expand an arbitrary CP map in
terms of Kraus operators $M_i$ \cite{kraus1}
\beq
\E{\r}=\sum_i M_i \r M^\dagger_i
\eeq
Of course, this operator-sum decomposition is
not unique,
but it can be shown
\cite{kraus1,kraus2} that all other
decomposition's have Kraus
operators $\N_j$
related by a unitary transformation
$\N_j=\M_i U_{ij}$.
The operator-sum decomposition may therefore be thought of as
being analogous to a density matrix.  In particular, it can be
shown \cite{schumacker-96} that for a given state $\r$, there
exists a diagonal representation, such that
\beq \tr M_\mu \r
M^\dagger_\nu= 0 \s \s for \  \mu\neq\nu \eeq
If $\rho$ is taken
to be the maximally entangled state $\singlet$, with the Kraus
operators acting on half of it, then one sees that the $M_\mu$ are
orthogonal under the trace inner product as with the orthogonal
unitaries (or the non-unitary set $A_\mu$) considered in the preceding section.  One therefore has
that
\beq \ket{\tilde{\psi}_\mu}=M_\mu\ket{\singlet} \eeq
are
orthogonal states (unnormalized).  The normalized states we call
$\ket{\psi_\mu}$.  After the CP map has acted on half the
$\singlet$, we are left with a density matrix given my \beq W=\sum
p_\mu \proj{\psi_\mu}. \eeq

One way to think of how the CP map arises
is to consider
a unitary which acts not only on $\r$, but also
on the
system plus an ancilla $\ket{0_C}$ (so-called Stinespring dilation)
namely
\beq
\E\rho=\tr_CU(\rho\proj{0_C})
\label{eq:stinespring}
\eeq
After considering such a global unitary, one can take the
ancilla to be with a third party (who we will call Charlie), who is considered to
be the source $\C$ of the CP map.

We then define the {\it entropy of a CP map} as
\beq
\so = -\sum p_\mu\log(p_\mu)
\label{eq:mapentropy}
\eeq
and show that it gives the rate at which one
can noiselessly
compress the informational content of the CP
map.
By information, we mean something analogous to the informational
content of a state under compression.  In the case of states, the compression
is done without knowing the ensemble, and after decompression, one
can verify that one faithfully obtained some series of states by
having the source read out each state that was sent.  One then
performs a measurement on the decompressed states to verify fidelity.

Here, in analogy with ensembles of states, we have choices of the Kraus
representation $M_i$.  We can therefore verify that all the
information of the CP map has been faithfully stored under the following test:
Charlie performs a measurement on the ancilla in an arbitrary basis.
We will see that choosing the basis is the equivalent of choosing some Kraus
representation
(like choosing the ensemble).
Charlie's result is in one to one correspondence with a particular
$M_i$, and we can verify that indeed this $M_i$ acted on our state.
We thus have a correspondence between the informational content of
states,
and that of operations.

To see this we consider  a measurement on the ancilla $\C$ 
in the basis $\ket{i_C}$ after the unitary $U$ of Eq. (\ref{eq:stinespring}) has
acted on the ancilla and $\singlet$.
This then selects the $M_i$ via
\beq
M_i\ket{\singlet}=\bra{i_C}U\ket{\singlet}\otimes\ket{0_C}\s .
\eeq
Therefore if given the value of $i$ from the source, one can verify
that $M_i$ did indeed act. Note that the particular form of
the $M_i$ is dependent on the state acted upon, although the CP
map itself is state independent.

That $\so$ qubits are necessary and sufficient to store this
information is straightforward.
The rate can be achieved for large $n$, simply
by having the source perform
each unitary on a maximally entangled state
$\singlet$, creating the
a density matrix given by
\beq
\rs = \sum p_\mu \proj{\sing{\mu}} \s .
\label{eq:storeddensity}
\eeq
From Shannon's noiseless coding theorem, the
state
with density matrix $\rs$ can be compressed
at a rate of $\so+\epsilon$ with $\epsilon$ as
small as desired
in the limit of large $n$.  The encoding
clearly preserves the informational content as described above.

That this rate is optimal, can be seen from the
fact that the encoding must work for all ensembles, and
in particular 
we could choose the
ensemble to be the set of orthogonal
operators $M_\mu$.  
A better compression rate would then
imply a violation of the Holevo bound.

A particular example of the above scheme are
CP maps which correspond to unitaries applied probabilistically.
We imagine that a sequence of
unitaries are performed
by a source $\C$, and that while we don't know
what
unitaries are being performed, nor from what
ensemble the
unitaries are being drawn from, we do know the
CP map that
the source performs.  Again, this is in analogy with
knowing the
density matrix of a source which emits states.
I.e. one images a sequence of
unitaries performed on the state, where
the unitaries are chosen from some
{\bf unknown} ensemble $\en=\{p_i,\U_i\}$,
(the $\U_i$ need not be orthogonal)
and we wish to compress a particular sequence
$\x$ of $n$ draws from this
ensemble.
All we are given is a Kraus representation
of the CP map.  Using the method above, the sequence
of $\U_i$ can be compressed at a rate $\so$, and one
can indeed verify whether any particular sequence $\x$ of
unitaries was performed.


One might hope that the information concerning
the sequence of
positive operators could be encoded and decoded in such
a way that a
recipient can act the map on an unknown
state given
after the encoding.
However, due to a no-go theorem for programmable unitary gates 
\cite{nielsonchuang-universal}, extended to the approximate 
case in \cite{VMC-2001},
this is
impossible for an arbitrary ensemble.  

This result is easily extended to the case of Kraus operators. 
Consider an unknown sequence
of Kraus operators $\x=M_1M_2M_3...M_n$ and similarly $\x'$,
and a distance measure $D(\x,\x')=\tr(|\x-\x'|)$.  A given
protocol aims to act $\x$ on an unknown set of states $\psi$,
generating the state $\Psi$  Call the error rate of a given protocol
$\epsilon=|\bra{\psi} M_1\ket{\psi}M_2\ket{\psi}M_3\ket{\psi}...|^2$

\begin{theorem}
Given $\x,\x'$ drawn from any
operator-sum decomposition of the CP map $\E{\cdot}$,
and any encoding $A(\E{\cdot})$
which maps sequences $\x,\x'$ to states  $\tau_x,\tau_x'$,
and decoding algorithm
$B(A(\E{\cdot}),\psi)$ which maps $\tau_x$
to $\ket{\Psi}$ 
close to $M_1\ket{\psi}M_2\ket{\psi}M_3\ket{\psi}...$
with error rate $\epsilon$.
Then if 
$\ket{\psi}$ is an arbitrary
unknown state chosen
after encoding, $\tr|\tau_x \tau_x'|\leq O(\sqrt{\epsilon})/D(\x,\x')$.
\end{theorem}

We refer the reader to \cite{VMC-2001} for the full proof,
and just give the no go theorem in the exact case,
%
%
using the fact that the
encoding must be unitary.  
The decoding takes as input, a state
$\ket{\psi}^{\otimes n}$, and
the encoding of the map realization $\tau_x$.  Let us
first take $\tau_x$
to be a pure state $\ket{x}$ (our proof will
extend to any
mixed state $\tau_x$ by the linearity of
quantum mechanics).
The decoding then takes this input and produces
the sequence
$\ket{Y}=M_1\ket{\psi}M_2\ket{\psi}M_3\ket{\psi}...$
and some
ancilla $\ket{\chi_x}$.  The ancilla cannot
depend on $\psi$ for coherence
to be preserved.
We can imagine the encoding being performed on
another sequence
$\x'$, encoded in $\ket{x'}$, and producing a
sequence
$\ket{Y'}=\M'_1\ket{\psi}M'_2\ket{\psi}M'_3\ket{\psi}...$,
and ancilla
 $\ket{\chi_{x'}}$.
Then, since the decoding must be unitary it
must preserve the
inner product of any two inputs
\beq
\bra{x}x'\ra =
\bra{\chi_x}\chi_{x'}\ra\bra{Y}Y'\ra  \s.
\label{eq:unitaritycond}
\eeq
Since neither $\bra{x}x'\ra$ nor $
\bra{\chi_x}\chi_{x'}\ra$
can depend on $\psi$ it follows that either
 $\bra{x}x'\ra =\bra{\chi_x}\chi_{x'}\ra=0$ or
$\bra{Y}Y'\ra$
cannot depend on $\psi$.  The latter can only
occur if
$\x=\x'$, therefore, if the encoding/decoding
is to work for different
possible inputs we require $\bra{x}x'\ra=0$.
I.e. an
orthogonal state must be chosen for each
possible sequence,
and the size of
the encoded state must then be as large as the
number of possible
sequences.  Since there are an arbitrarily
large number of possible
ensembles which implement a given CP map, it
follows that the
size of the encoded state must be infinite.
  In essence, the
size of the program grows with the size of the ensemble.

%
%

It is not clear if one can do better if the state is known
to the decoder.

It is perhaps amusing that there is an infinite discontinuity
which occurs if all $M_i$ are identical and perfect fidelity
is required.  One can imagine a
CP map which can be decomposed into two orthogonal unitaries $U_1$
and $U_2$ and that one is applied with probability $1-\epsilon$, and
the other with probability $\epsilon$.  
there is an infinite discontinuity in that
the  number of possible Kraus representations
goes from infinity to one.  The same discontinuity 
exists for ensembles of density matrices.
There is therefore potentially something special about 
pure states and pure unitaries.
This discontinuity only exists if one demands perfect fidelity of
the decoding, therefore it is unclear what the interpretation of this
observation is.
The above has the flavor of a phase-transition 
(c.f. \cite{dorit-phase,OHHphase}).

One can now ask whether one can perhaps act the compressed evolution
on an unknown state probabilistically.  Indeed, for the case of a 
stored phase gate of the form $U(\alpha)=exp(i\alpha\sigma_z)$ 
one can act the stored gate on an unknown state with probability
$1/2$ \cite{VMC-2001}.  We now generalize this to arbitrary unitaries
and Kraus operators.

Consider an unknown state $\psi$ and evolution $M_i$
stored in state $\psi_i$.  We then perform the unitary
\beq
V=\sum_\mu P_\mu M_\mu
\eeq
where $P_\mu$ are projector onto the orthogonal states $\psi_\mu$
which are eigenkets of $\rs$ defined via Eq. (\ref{eq:storeddensity}).
The stored evolution can be expanded in terms of the orthogonal set
of Kraus operators as
\beq
M_i=\sum c_{i\mu} M_\mu
\eeq
We then act $V$ on the stored evolution and the unknown state
\beq
V\ket{\psi_i}\otimes\ket{\psi}=\sum c_{i\mu} \psi_\mu \otimes \ket{M_\mu \psi}
\eeq
we then measure the state which was storing the unitary, in a basic
complementary to $\psi_\mu$.  For example, we can measure using projectors
onto $\psi_{\mu'}$ with $\bra{\psi_{\mu'}}\psi_\mu\ra=\pm 1/\sqrt{d}$.  
Then, with probability $1/d$ we will have succeeded in performing
the correct Kraus operator.
 
\section{Super-dense coding of unitaries}

The preceding section therefore gives an
informational interpretation of
evolutions.  In fact, one can regard the
entropy of Eq. (\ref{eq:mapentropy}) as
representing the maximum amount of information
that the
evolution can transfer from an environment or source
to a state.  This leads to a natural
generalization
of super-dense coding where the information
that is conveyed is
not classical bits, but rather, pure quantum
information.

One can imagine that two parties
(Alice and Bob) share a maximally entangled
state, and that
Alice has access to a source C of random unitaries
which acts on her
half of the singlet.  Alternatively, Alice
might apply
unitaries conditional on quantum states, or might
apply the unitaries herself according to some classical
probability distribution.
The action of the unitaries will produce a
sequence of maximally
entangled states shared between Alice and Bob.
After Alice sends her half of the singlet to
Bob, he will obtain all the quantum information
about the unitary.  Since the basis of qubit unitaries
is $2^n$ larger than the basis for qubit states,
this can therefore be viewed as a ``quantum'' version
of the classical communication sent in super-dense coding.
In the case of super-dense
coding, Alice chooses from $4$ orthogonal unitaries
and applies them to her half singlet and sends.
Here, one allows arbitrary superpositions of the orthogonal unitaries
to be applied.  What is more, the information that is sent
can be sent blindly.  Alice need not know which
unitaries are being applied by the source C.
If she first tried to know which unitaries were being applied
by the source, she would of course, destroy the quantum state.

An alternative generalization of super-dense coding has been
independently proposed  in \cite{harrow-superdense}.
There, it was shown that in large dimensions, using singlets
and shared randomness, Alice could send known quantum states
using only half as many qubits.

As with super-dense coding, the sending of the arbitrary unitary
is cryptographically secure, in that an eavesdropper, located
between Alice and Bob, obtains no information about which unitary
was applied (neither can Alice learn which unitary was applied,
as long as Bob holds the other half of the used singlet).  One may
therefore regard this as
a one way private quantum channel \cite{boykin-qe,
mosca-qe} which uses a resource of one ebit
per 2 qubits of sent information rather than 2
cbits for each qubit (although see the key-recycling results
of \cite{oh-recycling,debbie-qvc}),
or 2 ebits \cite{debbie-qvc} per qubit.

\section{Entanglement and Concentration of Unitaries}
Does the notion of entanglement extend to the
case of evolutions?
Consider a unitary interaction that acts on a
pair of systems.
Clearly, the combined evolution operator can be
expanded in terms
of the
unitary basis operators of each system
in the general form
\beq
U^{(I,II)} =\sum C_{\mu\nu}
U_{\mu}^{(I)}\otimes V_{\nu}^{(II)}
\eeq
where $U^{(I)}_\mu$ and $V^{(I)}_\nu$ are the
`local' orthogonal
unitary basis.
As consequence of (4) a measurement of say
system $I$,    will
lead
to a collapse of the sum to a single term.
Likewise the familiar entanglement bipartite
correlations are recovered for interactions.

In the sense of a passive transformation we can
re-express  the
general
state by performing the transformation
$A_\mu= \sum_\alpha
K_{\mu\alpha} U_{\alpha}$ and
$ B_\nu=\sum_\beta Y_{\nu\beta} V_\beta$,
such that $KCY=D$ is diagonalized in the new
orthogonal basis with eigenvalues $d_\mu$
Hence a Schmidt form can be written also for
unitary interactions
 \beq \label{schmidt}
\tilde U^{(I,II)} =\sum d_{\mu}
A_{\mu}^{(I)}\otimes \tilde
B_{\mu}^{(II)}
\eeq
The operators $A_{\mu}^{(I)}$ and $B_{\mu}^{(II)}$ in the above
decomposition are generally not
unitary, however they maintain orthogonality under the trace inner product.
Hence, we can apply the same procedure, described in section 2., to
measure which operator has acted on each side of the bipartite system.
The probability to find a certain operator $A_\mu$ (or $B_\mu$ if the measurement
takes place at side $II$) is then given by $|D_\mu|^2$. There is then
a one to one correlation between the results of the measurement of which
operator has acted on system $I$ and $II$.

We can now quantify the entanglement of the
interaction by computing the
 entropy of the probabilities, $-\sum |d_\mu|^2\log|d_\mu|^2$, in this diagonal
basis.
To justify this choice we demonstrate a
concentration procedure
for $n$ identical non-maximal bi-partite interactions.
We emphasis that we now consider a concentration process that
is independent of the nature of the state $\rho(I,II)$, on which the unitary
$U^{(I,II)}$ acts.

Suppose that we operate $n$ times the same bi-partite interaction
\beq\label{product}
\biggl[\alpha I^{(I)}\otimes I^{(II)} + \beta
\sigma_x^{(I)}\otimes \sigma^{(II)}_x
\biggr]^{\otimes n}
\eeq

We would like now to concentrate this "non-maximal interaction"
to a sum of terms with equal coefficients.
Recalling that in the state concentration scheme one employs a
collective measurement of the
 operator $J_z^{(I)}=\sum_i \sigma_{zi}^{I}$, we shall
now consider a measurement of the temporal collective correlation
$\Delta J^{(I)}_z(t_2,t_1)= J_z^{(I)}(t_2) - J_z^{(I)}(t_1)$ (more generally,
when we have a large number of terms one has to measure
more temporal correlations).
The equation $T: J(t_2,t_1)U_i= \lambda U_i$,
has solutions with eigenvalues $\lambda = (-n,-n+2, ....,n)$.
The relevant eigenoperators corresponding to $U_i$ have the structure of a sum
of  terms, where each of the terms is given by
 products of unit operators and Pauli operators.
The total number
of Pauli operators
is identical in all terms
and determined by the eigenvalue $\lambda$.
The coefficients of the terms need not be identical hence
$U$ above is generally degenerate.
Nevertheless, in our particular case,
a straightforward calculation shows that a measurement
of the operator $\Delta J_z(t_2,t_1)$, that may be performed on subsystem
$I$ or $II$, collapses (\ref{product}) to the operator
\beq\label{collapsed-product}
C_U =\biggr[ (I^{(I)}_1...I^{(I)}_k \sigma_{k+1}{(I)}....\sigma_n^{(I)})
(I^{(II)}_1....\sigma_n^{(II)})+....
\biggr]
\eeq
Notice that the terms in the square brackets above are now all
equally weighted, and their number is determined
 by the measurement outcome.
The probability to collapse into a particular value of operator is given
by $\alpha^{2k}\beta^{2(N-k)}$.
Therefore, in complete analogy to
the case of pure state concentration, the expected number of
equally weighted terms in $C_U$,  peaks in the limit of large
$n$ around $2^{nS(d_\mu)}$, where $S(d_\mu)$ is the Shannon entropy.
 Notice that in general the
operator $C_U$ is not unitary. Nevertheless, its entangling capability
power is equivalent to $n$ controlled-not interactions: it can convert
$n$ non-entangled pairs into a block with $2^{nS}$ equally weighted terms
which is maximally entangled. However, unlike the case of state concentration,
$C_U$ cannot be further factored by means of local
operations to a product of bi-partite maximally entangled unitaries.

The above result suggests a notion of
bi-partite {\em interaction-entanglement}, $S_U$,
which is a straightforward extension of
ordinary entanglement:
\beq
S_U= -\sum_\mu |d_\mu|^2\log |d_\mu|^2.
\eeq
This definition is with complete harmony with the entropy
defined previously for a CP map.
Therefore, given by a bi-partite unitary interaction,
the entanglement entropy of the interaction corresponds locally
to the entanglement of the locally generated CP map.
This can be seen by noticing that the operators $A_\mu$ ($B_\mu$) in the Schmidt
decomposition (\ref{schmidt})
are in fact then the same Kraus operators that appear
in the sum representation of the CP map which act on system $I$ ($II$).

The analog of a maximal
entangled state is in our case given by ${1\over \sqrt2} (I\otimes
I+i\sigma_x\otimes\sigma_x)$,
which is equivalent to a controlled-not (up to
additional local rotations).
We can now compare the proposed notion of
interaction-entanglement
with that
of  {\em entanglement capability} of an interaction
\cite{DVCLP-2001}.
The later is defined
by maximizing the amount of state-entanglement
that an interaction
produces by acting on a particularly chosen
state.
Clearly the two notions differ.
Interaction-entanglement
does not depend on the nature of the initial
state, while the entanglement capability clearly does.
Furthermore, in general the numerical value of
entanglement capability is larger than the interaction-entanglement
because one optimizes the entanglement gain over the initial states.
In contrast, the interaction entanglement, as well as the CP map entropy,
are independent of the entanglement content of the state.

\section{Conclusion}

The focus of this paper has been on giving an operational
interpretation
to the formal correspondence between operators and states,
and enlarging our view of the probabilistic interpretation
of quantum mechanics.
We have seen that one can treat operations in a similar
manner as one treats states.  By making a single measurement
one is able to say which operation acted on a state.
The probability of the result of this measurement is given
by a simple extension of the usual probability laws of
quantum mechanics, and is independent of the state that
gets acted on.  The results follow from the ordinary
laws of quantum mechanics, and yield interesting
interpretational issues. While the probabilistic interpretation and
collapse can be formulated in analogy to that of quantum states, it
remains to be seen to what extent can we truly interpret
the expansion of $U$ as a sum over unitary evolutions
as a quantum $superposition$ of evolutions.
One could object for instance to this interpretation by arguing
that while the final outcome of the measurement is indeed a
collapse to a single effective evolution $U_\mu$, the evolution of the
system  in between the
two intervention times, $t_1$ and $t_2$, is in fact not described
by the resulting $U_\mu$. Thus we have not collapsed to a single
unitary but only to an effectively equivalent unitary.  Such questions
do not bother us for the case of a single time measurement, and
it is not clear how to interpret such questions for the two-time measurements 
considered here.  

It also remains to be studied in what respects the probabilistic
interpretation of evolutions differs from the conventional interpretation.
One such important difference it that while non-orthogonal states
can not be distinguished
with certainty, non-orthogonal evolutions can. Understanding how this
can be incorporated in a rigorous probabilistic formalism is a potentially
rich area of research.  The information theoretic nature of evolutions has also
been explored, and we have given an information theoretic
interpretation to CP maps, using the idea of compression of their
informational contents.
For arbitrary realizations of a given CP map, we found that it was
possible to compress the map, and act it probabilistically on 
an unknown system.  It would be interesting to explore whether one
could act it on a known state given after compression. 
A generalization of superdense coding was also introduced.
With regard to our entanglement concentration scheme,
we have not yet touched on 
possible analogies for 
dilution for the case of interaction entanglement.
This leaves open the question whether the proposed measure of
interaction entanglement is a reversible quantity.

\begin{acknowledgments}
We are grateful to
Yakir Aharanov, Ignacio Cirac, Daniel Oi, and Lev Vaidman
for interesting discussions.
JO acknowledges the support of the Lady Davis
Trust, and ISF grant 129/00-1 as well as funding
by project PROSECCO (IST-2001-39227) of the IST-FET program of the EC and 
a grant from the Cambridge-MIT Institute.
BR acknowledges the support of ISF grant 62/01-1.  This research
was conducted during the Banasque session on Quantum Information
and Communication, 2003, and we thank the town and the organizers
for their hospitality.

\end{acknowledgments}

\section{Appendix}
In this appendix we demonstrate our probabilistic interpretation
and measurement scheme for the general d-dimensional case.
Let us denote the orthogonal
basis as $U_{\mu\nu}$ where he two indices take the values
$\mu,\nu = 0,...,d-1$. Then
\beq
U_{\mu\nu}= U_0 \sigma_{\mu\nu}
\eeq
where $\sigma_{\mu\nu}$ are $d^2$ traceless orthogonal unitary
operators.
We will consider first the simple case where
\beq
\sigma_{\mu\nu}= (Z)^\mu (X)^\nu
\eeq
where the operators \cite{schwinger}
\beq
Z= \sum_{j=0}^{N-1} \zeta^j|j\ra\la j|
\eeq
and
\beq
X= \sum_{j=0}^{N-1} |(j+1)mod\ N\ra\la j|
\eeq
are operators satisfying  $Z^N=X^N=1$ and $ZX=\zeta XZ$, where
$\zeta=  \exp(2\pi i/N)$.
We notice that for  $N=2$,
$Z\to \sigma_z$ and $X\to \sigma_x$ and regain our previous
construction
using Pauli operators.
Thus $Z$ and $X$  play the role of generalized `phase flip'
and `bit flip' operators.

The extension of the eigenoperators is then given by
\beqa A_X(t_2;t_1)= (U_0XU_0)_{t_2}(X)_{t_1}+ h.c \\
A_Z(t_2;t_1) =(U_0ZU_0)_{t_2}(Z)_{t_1}+h.c.
\eeqa
As we will shortly see, $A_X$ and $A_Z$ ascribe
the value of the first and second indexes of a single element of
the unitary basis $U_{\mu\nu}$.

To perform the measurement we employ now a pair of N-level
ancillary systems
in the initial state $\sum  |\tilde\alpha\ra\sum|\beta\ra$.
The interaction operator
can be expressed as
\beq
V_Z  = \sum_{\alpha=0}^{N-1} |\alpha\ra\la \alpha| Z^\alpha
\eeq
and similarly we define $\tilde V_X$.
The sequence of interaction at $t_1$, free evolution, and
interaction at $t_2$
then reads
\beq
(\tilde V_X V_Z) (\sum C_{\mu\nu}\sigma_{\mu\nu}) (\tilde V_X V_Z)
\eeq
where for the simplicity of presentation we have  dropped the
$U_0$ factor.
Acting on the total state we obtain
\beq
\sum_{\mu\nu} C_{\mu\nu}\biggl[\sum_\alpha X^\alpha
\sigma_{\mu\nu}X^\alpha \sigma^\dagger_{\mu\nu}|\tilde\alpha\ra
\sum_\beta Z^\beta \sigma_{\mu\nu}Z^\beta \sigma^\dagger_{\mu\nu}
|\beta\ra\biggr]\sigma_{\mu\nu}|\psi\ra
\eeq
The main point is that the operators
$X^\alpha \sigma_{\mu\nu}X^\alpha \sigma^\dagger_{\mu\nu}$,
and $ Z^\beta \sigma_{\mu\nu}Z^\beta \sigma^\dagger_{\mu\nu}$ are
constants
of motion.
Using the commutation relation of $X$ and $Z$ we finally get
\beq
\sum_{\mu\nu} C_{\mu\nu}\biggl[\sum_\alpha
\zeta^{\alpha\mu}|\tilde\alpha\ra
\sum_\beta \zeta^{\beta\mu}|\beta\ra\biggr]\sigma_{\mu\nu}|\psi\ra
\eeq
\beq
\equiv\sum_{\mu\nu}C_{\mu_\nu}|\tilde\phi_\mu\ra|\phi_\nu\ra
\sigma_{\mu\nu}|\psi\ra
\eeq
where the ancilla states $\tilde \phi_\mu$ and $\phi_\nu$ are
orthogonal,
hence a measurement at $t=t_2$ will indeed collapse the sum to a
single term
with a probability $|C_{\mu\nu}|^2$, and leave only the unitary
$\sigma_{\mu\nu}$.

More generally, it is known that for  $d\ge3$ there
are different inequivalent unitary basis \cite{werner-alltel}. However
there
exists a one-to-one correspondence between the unitary
basis and the basis of maximally entangled states
\cite{werner2}.
Since for the latter we can always identify an observable which
distinguishes between the basis elements,  a corresponding
observable can be constructed for an arbitrary unitary basis.

We finally note, that the generalization of the particular
construction above
to the case of a continuous Hilbert space is
straightforward. In this case $\sigma_{\mu\nu} \to \sigma_{x_0p_0}
=T_{x_0}T_{p_0}$, where
\beqa
T_{x_0}&=&\int dx  |x+x_0\ra\la x| \\
T_{p_0}&=&\int dx e^{ixp_0}|x\ra\la x|
\eeqa
The general set of orthogonal unitary operators is then
$U_{x_0p_0}= U_0(x,p)
\sigma_{x_0p_0}$, where $x_0$ and $p_0$ are continuous real
numbers.
\beq
C_{x_0p_0} = \int dx e^{ix p_0} \la x+x_0|U|x\ra
\eeq
The amplitude of  a basis element for a general $U$
has then a form similar to the Wigner-distribution


\begin{thebibliography}{29}
\expandafter\ifx\csname natexlab\endcsname\relax\def\natexlab#1{#1}\fi
\expandafter\ifx\csname bibnamefont\endcsname\relax
  \def\bibnamefont#1{#1}\fi
\expandafter\ifx\csname bibfnamefont\endcsname\relax
  \def\bibfnamefont#1{#1}\fi
\expandafter\ifx\csname citenamefont\endcsname\relax
  \def\citenamefont#1{#1}\fi
\expandafter\ifx\csname url\endcsname\relax
  \def\url#1{\texttt{#1}}\fi
\expandafter\ifx\csname urlprefix\endcsname\relax\def\urlprefix{URL }\fi
\providecommand{\bibinfo}[2]{#2}
\providecommand{\eprint}[2][]{\url{#2}}

\bibitem[{\citenamefont{Jamiolkowski}(1972)}]{jam-72}
\bibinfo{author}{\bibfnamefont{A.}~\bibnamefont{Jamiolkowski}},
  \bibinfo{journal}{Rep. Math. Phys.} \textbf{\bibinfo{volume}{3}},
  \bibinfo{pages}{275} (\bibinfo{year}{1972}).

\bibitem[{\citenamefont{Helkwig and Kraus}(1970)}]{kraus1}
\bibinfo{author}{\bibfnamefont{K.}~\bibnamefont{Helkwig}} \bibnamefont{and}
  \bibinfo{author}{\bibfnamefont{K.}~\bibnamefont{Kraus}},
  \bibinfo{journal}{Commun. Math. Phys.} \textbf{\bibinfo{volume}{16}},
  \bibinfo{pages}{142} (\bibinfo{year}{1970}).

\bibitem[{\citenamefont{Dur and Cirac}(2001)}]{durcirac-ops}
\bibinfo{author}{\bibfnamefont{W.}~\bibnamefont{Dur}} \bibnamefont{and}
  \bibinfo{author}{\bibfnamefont{I.}~\bibnamefont{Cirac}},
  \bibinfo{journal}{Phys. Rev. A} \textbf{\bibinfo{volume}{64}},
  \bibinfo{pages}{012317} (\bibinfo{year}{2001}).

\bibitem[{\citenamefont{Vidal et~al.}(2002)\citenamefont{Vidal, Masanes, and
  Cirac}}]{VMC-2001}
\bibinfo{author}{\bibfnamefont{G.}~\bibnamefont{Vidal}},
  \bibinfo{author}{\bibfnamefont{L.}~\bibnamefont{Masanes}}, \bibnamefont{and}
  \bibinfo{author}{\bibfnamefont{J.}~\bibnamefont{Cirac}},
  \bibinfo{journal}{Phys. Rev. Lett.} \textbf{\bibinfo{volume}{88}},
  \bibinfo{pages}{047905} (\bibinfo{year}{2002}), \eprint{quant-ph/0102037}.

\bibitem[{\citenamefont{Nielson and Chuang}(1997)}]{nielsonchuang-universal}
\bibinfo{author}{\bibfnamefont{M.}~\bibnamefont{Nielson}} \bibnamefont{and}
  \bibinfo{author}{\bibfnamefont{I.}~\bibnamefont{Chuang}},
  \bibinfo{journal}{Phys. Rev. Lett.} \textbf{\bibinfo{volume}{79}},
  \bibinfo{pages}{321} (\bibinfo{year}{1997}), \eprint{quant/ph-9703032}.

\bibitem[{\citenamefont{Collins et~al.}(2001)\citenamefont{Collins, Linden, and
  Popescu}}]{clp-2000}
\bibinfo{author}{\bibfnamefont{D.}~\bibnamefont{Collins}},
  \bibinfo{author}{\bibfnamefont{N.}~\bibnamefont{Linden}}, \bibnamefont{and}
  \bibinfo{author}{\bibfnamefont{S.}~\bibnamefont{Popescu}},
  \bibinfo{journal}{Phys. Rev. A} \textbf{\bibinfo{volume}{64}},
  \bibinfo{pages}{032302} (\bibinfo{year}{2001}), \eprint{quant-ph/0005102}.

\bibitem[{\citenamefont{Cirac et~al.}(2001)\citenamefont{Cirac, Dur, Kraus, and
  Lewnstein}}]{cdkl-2000}
\bibinfo{author}{\bibfnamefont{J.~I.} \bibnamefont{Cirac}},
  \bibinfo{author}{\bibfnamefont{W.}~\bibnamefont{Dur}},
  \bibinfo{author}{\bibfnamefont{B.}~\bibnamefont{Kraus}}, \bibnamefont{and}
  \bibinfo{author}{\bibfnamefont{M.}~\bibnamefont{Lewnstein}},
  \bibinfo{journal}{Phys. Rev. Lett.} \textbf{\bibinfo{volume}{86}},
  \bibinfo{pages}{2001} (\bibinfo{year}{2001}).

\bibitem[{\citenamefont{Zanardi et~al.}(2000)\citenamefont{Zanardi, Zalka, and
  Faoro}}]{zanardi-etal}
\bibinfo{author}{\bibfnamefont{P.}~\bibnamefont{Zanardi}},
  \bibinfo{author}{\bibfnamefont{C.}~\bibnamefont{Zalka}}, \bibnamefont{and}
  \bibinfo{author}{\bibfnamefont{L.}~\bibnamefont{Faoro}},
  \bibinfo{journal}{Phys. Rev. A} \textbf{\bibinfo{volume}{62}},
  \bibinfo{pages}{030301(R)} (\bibinfo{year}{2000}), \eprint{quant-ph/0005031}.

\bibitem[{\citenamefont{Zanardi}(2001)}]{zanardi-2001}
\bibinfo{author}{\bibfnamefont{P.}~\bibnamefont{Zanardi}},
  \bibinfo{journal}{Phys. Rev. A} \textbf{\bibinfo{volume}{63}},
  \bibinfo{pages}{040304(R)} (\bibinfo{year}{2001}).

\bibitem[{\citenamefont{Wang and Zanardi}(2002)}]{wang-zanardi}
\bibinfo{author}{\bibfnamefont{X.}~\bibnamefont{Wang}} \bibnamefont{and}
  \bibinfo{author}{\bibfnamefont{P.}~\bibnamefont{Zanardi}},
  \bibinfo{journal}{Phys. Rev. A} \textbf{\bibinfo{volume}{66}},
  \bibinfo{pages}{044303} (\bibinfo{year}{2002}), \eprint{0207007}.

\bibitem[{\citenamefont{Werner}()}]{werner-alltel}
\bibinfo{author}{\bibfnamefont{R.}~\bibnamefont{Werner}},
  \eprint{quant-ph/0003070}.

\bibitem[{\citenamefont{Acin}(2001)}]{acin-unitaries}
\bibinfo{author}{\bibfnamefont{A.}~\bibnamefont{Acin}}, \bibinfo{journal}{Phys.
  Rev. Lett.} \textbf{\bibinfo{volume}{87}}, \bibinfo{pages}{177901}
  (\bibinfo{year}{2001}), \eprint{quant-ph/0102064}.

\bibitem[{\citenamefont{Aharonov and Albert}(1984)}]{aharonov-albert}
\bibinfo{author}{\bibfnamefont{Y.}~\bibnamefont{Aharonov}} \bibnamefont{and}
  \bibinfo{author}{\bibfnamefont{D.}~\bibnamefont{Albert}},
  \bibinfo{journal}{Phys. Rev. D} \textbf{\bibinfo{volume}{29}},
  \bibinfo{pages}{223} (\bibinfo{year}{1984}).

\bibitem[{\citenamefont{Vaidman}(1994)}]{vaidman-teleportation}
\bibinfo{author}{\bibfnamefont{L.}~\bibnamefont{Vaidman}},
  \bibinfo{journal}{Phys. Rev. A} \textbf{\bibinfo{volume}{49}},
  \bibinfo{pages}{1473} (\bibinfo{year}{1994}).

\bibitem[{\citenamefont{Schumacher}(1995)}]{Schumacher1995}
\bibinfo{author}{\bibfnamefont{B.}~\bibnamefont{Schumacher}},
  \bibinfo{journal}{Phys. Rev. A} \textbf{\bibinfo{volume}{51}},
  \bibinfo{pages}{2738} (\bibinfo{year}{1995}).

\bibitem[{\citenamefont{Petz and Mosonyi}()}]{PetzMosonyi}
\bibinfo{author}{\bibfnamefont{D.}~\bibnamefont{Petz}} \bibnamefont{and}
  \bibinfo{author}{\bibfnamefont{M.}~\bibnamefont{Mosonyi}},
  \eprint{quant-ph/9912103}.

\bibitem[{\citenamefont{Hiai and Petz}(1991)}]{HiaiPetz91}
\bibinfo{author}{\bibfnamefont{F.}~\bibnamefont{Hiai}} \bibnamefont{and}
  \bibinfo{author}{\bibfnamefont{D.}~\bibnamefont{Petz}},
  \bibinfo{journal}{Commun. Math. Phys.} \textbf{\bibinfo{volume}{143}},
  \bibinfo{pages}{99} (\bibinfo{year}{1991}).

\bibitem[{\citenamefont{Schumacker}(1996)}]{schumacker-96}
\bibinfo{author}{\bibfnamefont{B.}~\bibnamefont{Schumacker}},
  \bibinfo{journal}{Phys. Rev. A} \textbf{\bibinfo{volume}{54}},
  \bibinfo{pages}{4707} (\bibinfo{year}{1996}).

\bibitem[{\citenamefont{Kraus}(1983)}]{kraus2}
\bibinfo{author}{\bibfnamefont{K.}~\bibnamefont{Kraus}},
  \emph{\bibinfo{title}{States, Effects and Operations: Fundemental Notions in
  Quantum Theory}} (\bibinfo{publisher}{Springer-Verlag, Berlin},
  \bibinfo{year}{1983}).

\bibitem[{\citenamefont{Aharonov}()}]{dorit-phase}
\bibinfo{author}{\bibfnamefont{D.}~\bibnamefont{Aharonov}},
  \eprint{quant-ph/9910081}.

\bibitem[{\citenamefont{Oppenheim et~al.}(2003)\citenamefont{Oppenheim,
  Horodecki, and Horodecki}}]{OHHphase}
\bibinfo{author}{\bibfnamefont{J.}~\bibnamefont{Oppenheim}},
  \bibinfo{author}{\bibfnamefont{M.}~\bibnamefont{Horodecki}},
  \bibnamefont{and}
  \bibinfo{author}{\bibfnamefont{R.}~\bibnamefont{Horodecki}},
  \bibinfo{journal}{Phys. Rev. Lett.} \textbf{\bibinfo{volume}{90}},
  \bibinfo{pages}{010404} (\bibinfo{year}{2003}), \eprint{quant-ph/0207169}.

\bibitem[{\citenamefont{Harrow et~al.}()\citenamefont{Harrow, Hayden, and
  Leung}}]{harrow-superdense}
\bibinfo{author}{\bibfnamefont{A.}~\bibnamefont{Harrow}},
  \bibinfo{author}{\bibfnamefont{P.}~\bibnamefont{Hayden}}, \bibnamefont{and}
  \bibinfo{author}{\bibfnamefont{D.}~\bibnamefont{Leung}},
  \eprint{quant-ph/0307221}.

\bibitem[{\citenamefont{Boykin and Roychowdhury}()}]{boykin-qe}
\bibinfo{author}{\bibfnamefont{P.}~\bibnamefont{Boykin}} \bibnamefont{and}
  \bibinfo{author}{\bibfnamefont{V.}~\bibnamefont{Roychowdhury}},
  \eprint{quant-ph/0003059}.

\bibitem[{\citenamefont{Mosca et~al.}(2000)\citenamefont{Mosca, Tapp, and
  de~Wolf}}]{mosca-qe}
\bibinfo{author}{\bibfnamefont{M.}~\bibnamefont{Mosca}},
  \bibinfo{author}{\bibfnamefont{A.}~\bibnamefont{Tapp}}, \bibnamefont{and}
  \bibinfo{author}{\bibfnamefont{R.}~\bibnamefont{de~Wolf}}, in
  \emph{\bibinfo{booktitle}{Proceedings of the 41st Annual Symposium on
  Foundations of Computer Science}} (\bibinfo{publisher}{IEEE Computer Society
  Press}, \bibinfo{year}{2000}), p. \bibinfo{pages}{547},
  \eprint{quant-ph/0003101}.

\bibitem[{\citenamefont{Oppenheim and Horodecki}()}]{oh-recycling}
\bibinfo{author}{\bibfnamefont{J.}~\bibnamefont{Oppenheim}} \bibnamefont{and}
  \bibinfo{author}{\bibfnamefont{M.}~\bibnamefont{Horodecki}},
  \eprint{quant-ph/0306161}.

\bibitem[{\citenamefont{Leung}(2001)}]{debbie-qvc}
\bibinfo{author}{\bibfnamefont{D.}~\bibnamefont{Leung}}, \bibinfo{journal}{QIC}
  \textbf{\bibinfo{volume}{2}}, \bibinfo{pages}{13} (\bibinfo{year}{2001}).

\bibitem[{\citenamefont{Dur et~al.}(2001)\citenamefont{Dur, Vidal, Cirac,
  Linden, and Popescu}}]{DVCLP-2001}
\bibinfo{author}{\bibfnamefont{W.}~\bibnamefont{Dur}},
  \bibinfo{author}{\bibfnamefont{G.}~\bibnamefont{Vidal}},
  \bibinfo{author}{\bibfnamefont{J.~I.} \bibnamefont{Cirac}},
  \bibinfo{author}{\bibfnamefont{N.}~\bibnamefont{Linden}}, \bibnamefont{and}
  \bibinfo{author}{\bibfnamefont{S.}~\bibnamefont{Popescu}},
  \bibinfo{journal}{Phys. Rev. Lett.} \textbf{\bibinfo{volume}{87}},
  \bibinfo{pages}{137901} (\bibinfo{year}{2001}), \eprint{quant-ph/00060034}.

\bibitem[{\citenamefont{Schwinger}(1960)}]{schwinger}
\bibinfo{author}{\bibfnamefont{J.}~\bibnamefont{Schwinger}},
  \bibinfo{journal}{Proc. Nat. Acad. Sci.} \textbf{\bibinfo{volume}{46}},
  \bibinfo{pages}{570} (\bibinfo{year}{1960}).

\bibitem[{\citenamefont{Vollbrecht and Werner}()}]{werner2}
\bibinfo{author}{\bibfnamefont{K.}~\bibnamefont{Vollbrecht}} \bibnamefont{and}
  \bibinfo{author}{\bibfnamefont{R.}~\bibnamefont{Werner}},
  \eprint{quant-ph/9910064}.

\end{thebibliography}

\end{document}